\documentclass[fleqn,usenatbib,useAMS]{mnras}
\usepackage[T1]{fontenc}
\usepackage{ae,aecompl}

\usepackage{graphicx}
\usepackage{color}
\usepackage{natbib}
\usepackage{mathtools}
\usepackage{amsmath}	
\usepackage{amssymb}	

\newcommand{\delchisq}{$\Delta \chi^2$}
\newcommand{\chisq}{$\chi^2$}
\newcommand{\tbary}{$t_{\mathrm{bary}}$}
\newcommand{\kepler}{\emph{Kepler}}
\renewcommand{\vec}[1]{\protect\overrightarrow{#1}}
\newcommand\Tstrut{\rule{0pt}{2.6ex}}         
\newcommand\Bstrut{\rule[-1.6ex]{0pt}{0pt}}   
\newcommand{\edit}[1]{{#1}}

\title[An Automated CBP Detection Algorithm]{An Automated Method to Detect Transiting \edit{Circumbinary Planets}}

\author[Windemuth et al.]{
Diana Windemuth$^{1}$\thanks{E-mail: windemut@uw.edu},
Eric Agol$^{1}$,
Josh Carter$^{2}$,
Eric B. Ford$^{3,4,5}$,
\newauthor\ Nader Haghighipour$^{6}$,
Jerome A. Orosz$^{7}$,
and William F. Welsh$^{7}$
\\
$^{1}$Department of Astronomy, University of Washington, Seattle, WA 98195\\
$^{2}$Citadel LLC, 131 S Dearborn St, Chicago, IL 60603\\
$^{3}$Department of Astronomy \& Astrophysics, 525 Davey Lab, The Pennsylvania State University, University Park, PA 16802\\
$^{4}$Center for Exoplanets \& Habitable Worlds, 525 Davey Lab, The Pennsylvania State University, University Park, PA 16802\\
$^{5}$Institute for CyberScience, The Pennsylvania State University, University Park, PA 16802\\
$^{6}$Institute for Astronomy, University of Hawaii, Honolulu, HI 96822\\
$^{7}$Department of Astronomy, San Diego State University, San Diego, CA 92182
}

\date{Accepted XXX. Received YYY; in original form ZZZ}

\pubyear{2019}

\begin{document}
\label{firstpage}
\pagerange{\pageref{firstpage}--\pageref{lastpage}}
\maketitle

\begin{abstract}
To date a dozen transiting ``Tatooines" or circumbinary planets (CBPs) have been discovered, by eye, in the data from the \kepler\ mission; by contrast, thousands of confirmed circumstellar planets orbiting around single stars have been detected using automated algorithms. Automated detection of CBPs is challenging because their transits are strongly aperiodic with irregular profiles. Here, we describe an efficient and automated technique for detecting circumbinary planets that transit their binary hosts in \kepler\ light curves. Our method accounts for large transit timing and duration variations (TTVs and TDVs), induced by binary reflex motion, in two ways: 1) We directly correct for large-scale TTVs and TDVs in the light curves by using Keplerian models to approximate binary and CBP orbits; and 2) We allow additional aperiodicities on the corrected light curves by employing the Quasi-periodic Automated Transit Search algorithm (QATS). We demonstrate that our method dramatically improves detection significance using simulated data and two previously identified CBP systems, \kepler-35 and \kepler-64. 
\end{abstract}


\section{Introduction}
\label{sec:intro}
The discovery rate of extrasolar planets orbiting around single stars has increased exponentially in the past two decades. This was largely enabled by the \kepler\ mission and the transit technique in synergy with ground-based radial velocity surveys. In stark contrast to the thousands of confirmed and candidate \emph{circumstellar} planets (CSPs), spanning Earth to Jovian in size, only \edit{11 published} transiting \emph{circumbinary} planets (CBPs) have been discovered in the \kepler\ data\footnote{The published CBPs are: Kep-16\emph{b} \citep{Doyle2011}; Kep-34\emph{b} \& Kep-35\emph{b} \citep{Welsh2012}; Kep-38\emph{b} \citep{Orosz2012-kep38}; PH1/Kep-64\emph{b} \citep{Schwamb2013, Kostov2013}; Kep-47\emph{b}, \emph{c}, and \emph{d} \citep{Orosz2012-kep47, Kostov2013, Hinse2015, Welsh2015-kep47, Orosz2019}; Kep-413\emph{b} \citep{Kostov2014}; Kep-453\emph{b} \citep{Welsh2015-kep453}; Kep-1647\emph{b} \citep{Kostov2016}. A 12th transiting circumbinary planet, KIC 10753734 has been reported but not yet published \citep{Orosz2016}.}. The transits of these CBPs across their binary hosts are sufficiently deep that they were identified by eye, while transit detections of planets around single stars were facilitated by automated detection pipelines \citep[e.g.][]{Petigura2013,Foreman-Mackey2016,Thompson2018}. 

Here, we present an automated technique to detect transiting CBPs. Our method \edit{assumes a single test planet in a circular and co-planar orbit around the binary, and} reconciles changes in transit timing and shapes in CBP systems in two ways: 1) We correct for circular and eccentric binary orbits using Keplerian models, creating a regularized light curve in which the CBP transit shape and timing are more uniform. 2) We apply the Quasi-periodic Automated Transit Search (QATS; \citealp{Carter2013}) algorithm on the regularized light curves, which allows for additional timing deviations due to \edit{inaccuracies in the binary model (e.g., mass ratios) and CBP model (e.g., planet eccentricity, perturbations of the planet's orbit by the binary or other planetary companions).}
This two-fold TTV treatment increases detection significance and decreases false alarm probability, and enables a systematic search for smaller, perhaps terrestrial, circumbinary worlds.  

Automatic transit detection algorithms not only enable larger-scale discoveries, they also permit statistical testing of detector efficiency, bias, and completeness. These quantities are imperative to robustly infer the underlying physical distribution of exoplanets, e.g., occurrence rates and system architectures  \citep{Batalha2013,Petigura2013,Foreman-Mackey2014,Ballard2016,Hsu2019}. A successful transit detection algorithm is optimized for speed and detection significance, and is typically conditioned on periodic signals. Notable exceptions to strict periodicity include the QATS algorithm, which allows a user-specified TTV window and have been applied to multi-planet systems around single stars, and algorithms tailored to detect transiting CBPs \citep{Jenkins1996, Ofir2008, Armstrong2014, Klagyivik2017}. 

Automated detection of CBP transits poses unique challenges and have lagged behind their CSP counterparts. This is because time-varying radiative and dynamical effects from the binary make transit pulses irregular in the time-series data. In particular, the reflex motion of the binary about the barycenter leads to ``roaming conjunctions" and induces large transit timing and duration variations (TTVs and TDVs), as well as transit depth variations in binary systems with unequal surface brightness. In fortuitous configurations, multiple transit events across the same or both stellar disks may occur near one syzygy (conjunction of observer-planet-stellar component). There may be additional variations in transit shape and timing due to precession of the CBP orbit \citep{Martin2015-transitprob}, which can cause transits to disappear and reappear over time, and due to second-order dynamical effects, e.g., perturbations from additional planets in the system. The latter is negligible compared to the geometric effect due to binary motion. Binary-induced TTVs can be of order days to even weeks (e.g., Kepler-16b; \citealp{Doyle2011}), orders of magnitude larger than TTVs due to perturbing companions in multi-planet systems around single stars \citep{Agol2005}. 

Previous work on transiting CBP detection addressed TTVs either by using a modified Box-fitting Least-Square (BLS; \citealp{Kovacs2002}) with a large, variable width \citep{Armstrong2014,Klagyivik2017}, or by directly correcting for binary motion and then applying a periodic search \citep{Jenkins1996,Ofir2008}. In particular, \cite{Armstrong2014} applied their algorithm to non-contact \kepler\ eclipsing binaries (EBs) to estimate the occurrence rate of CBPs, using a constant transit duration and analytic approximations \citep{Armstrong2013} to inform the maximum variable BLS window for each binary. \cite{Klagyivik2017} used a similar method to search for transiting CBPs in CoRoT data, but allowed the depth and duration of model transits to vary. Both studies, however, used a large TTV window in which to look for transits, which introduces a higher false-alarm probability.

\cite{Jenkins1996} and \cite{Ofir2008} proposed to directly correct for binary motion by solving the binary orbit to predict planet-binary positions. However, they required precise periodicity of planet transit times after an exact correction for the binary orbit, neither of which will be realized in practice. 

Here, we introduce a hybrid technique that corrects for geometric transit effects due to binary motion with a physical circumbinary orbit model {\it and} relaxes the subsequent assumption of strict periodicity using the QATS algorithm. In doing so, this paper is organized as follows. In Section \ref{sec:method}, we describe the geometry of CBP transits and detail our detection technique, including the physical binary and CBP models used for transit regularization. In Section \ref{sec:application}, we apply our method to simulated data, and also demonstrate its robustness on two known \kepler\ CBPs, \kepler-35\emph{b} \citep{Welsh2012} and \kepler-64\emph{b}/PH1\emph{b} \citep{Kostov2013,Schwamb2013}, which were tested with existing automated algorithms \citep{Armstrong2014, Klagyivik2017}. A detailed comparison of CBP detection techniques, such as one for generic transit detection algorithms by \cite{Tingley2004}, is beyond the scope of this paper. In Section \ref{sec:discussion}, we discuss the improvements on detection significance and comment on the limits of our detection technique. We conclude this study and summarize key takeaways in Section \ref{sec:conclusions}.

\section{Methods}
\label{sec:method}

Our detection technique, dubbed ``QATS-EB," differs from traditional transit search methods and addresses the aperiodic transits of circumbinary planets in a two-component approach. First, we use a semi-analytic CBP model to account for the influence of the binary hosts, which induce large-scale TTVs and TDVs on the planet. Second, we utilize the QATS algorithm \citep{Carter2013}, which relaxes strict periodicity in the transit search window with a user-specified inter-transit duration, i.e., allows room for error in timing predictions due to model inaccuracies. These two improvements work in conjunction to boost CBP detection signal-to-noise ratio (S/N) and decrease the false alarm probability. 

\begin{figure*}
\includegraphics[width=1.0\textwidth]{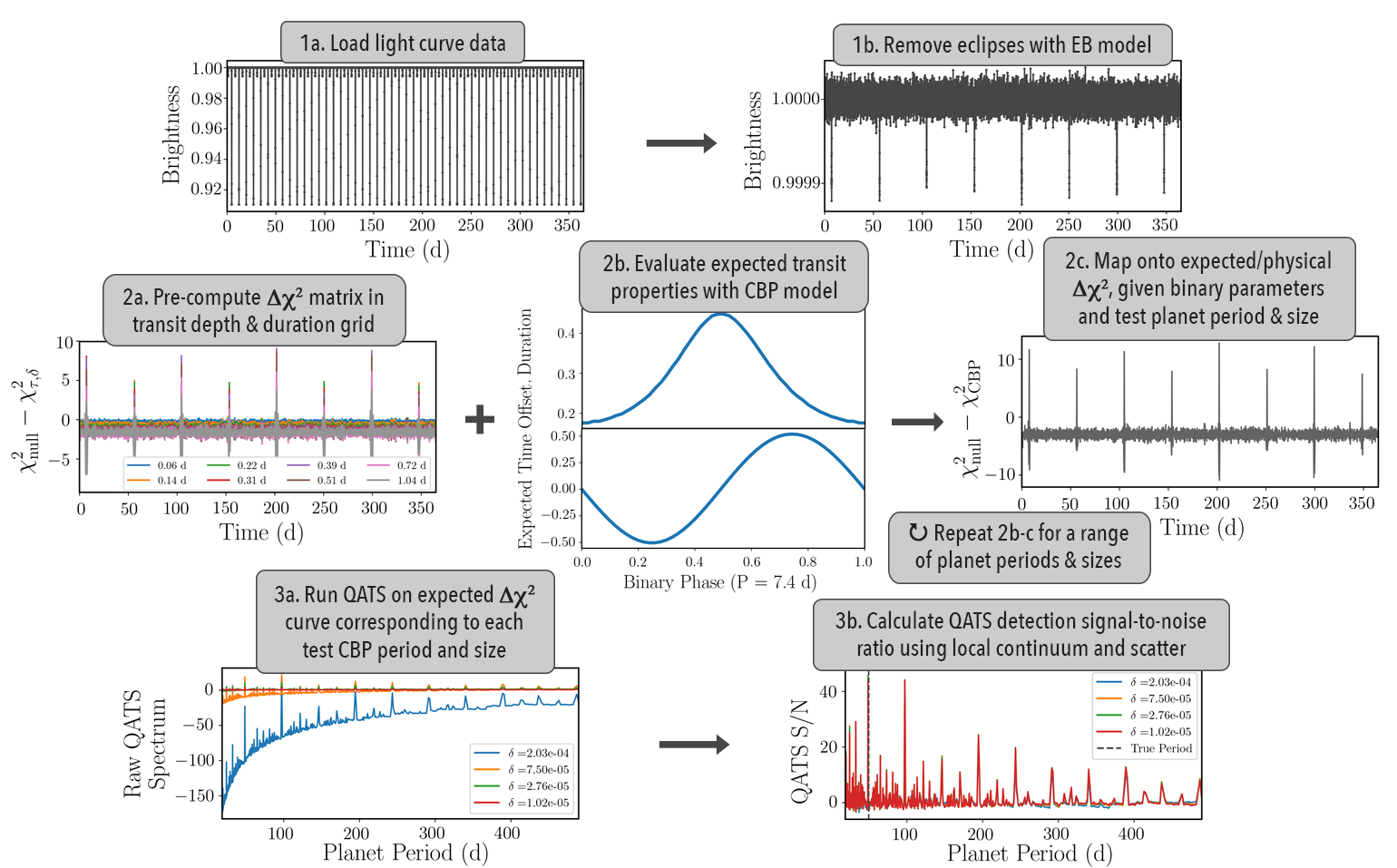}
\caption{A schematic diagram depicting procedures in the QATS-EB algorithm to detect circumbinary planets. (1) Use an EB model \citep{Windemuth2019} to remove eclipses in the light curve. (2a) Test EB-removed light curve for the presence of a transit with fixed depth and duration at each light curve cadence, creating a corresponding \delchisq-curve. (2b) Using a physical CBP model, compute the expected transit properties (depth, duration, barycentric timing offset) as a function of binary phase (or time), for a specified test CBP planet period and size. (2c) Interpolate the expected, regularized \delchisq-curve using (2a) and (2b). (3a) Run QATS on the expected \delchisq-curve, and loop through different test periods and sizes to build up maximum CBP likelihood as a function of period and size (QATS spectrum). (3c) Normalize each QATS spectral continuum to produce QATS signal-to-noise for detection statistic. See text in \S\ref{sec:method} for details.}
\label{fig:cbp_methods} 
\end{figure*}
Figure \ref{fig:cbp_methods} illustrates the overall QATS-EB procedure, which we preview in brief here. First, our method uses an EB model \citep{Windemuth2019} to solve for binary parameters and remove eclipses from the light curve (panels 1a--1b). Then, it performs, at each observed cadence, a likelihood ratio (chi-square difference \delchisq) test for a null hypothesis against a CBP transit across one of the binary stars, with predicted depth, duration, timing given by binary parameters and test planet period and size (panels 2a--2c). Finally, at each proposed planet parameters, it runs a quasi-periodic search on the time-series likelihood ratios and computes the detection S/N (panels 3a--3b). 

QATS-EB can search for CBP transits across the primary and secondary stars, in series. The predicted transit depth, duration, and timings are different for each stellar component, and the primary star is typically more luminous than the secondary. Thus, one can apply the method to search for CBP transits across the primary, then mask the detected transits and perform a subsequent, self-consistent search for transits across the secondary. Unless otherwise specified, we assume transits are across the primary star for the remainder of this paper. 

In the subsections to follow, we give an overview of definitions and the CBP transit geometry (\S\ref{subsec:definitions}) and provide detailed descriptions of the auomated CBP transit detection procedure (\S\ref{subsec:cbp_detection}). 

\subsection{Definitions \& Geometry of CBP Transits}
\label{subsec:definitions}
Throughout the paper, we use the following convention to describe the components of a circumbinary system. We use ``eclipses" to specify when the stellar components pass in front of each other, and use ``transits" when the planet passes in front of one of the stars.
\begin{figure*}
\includegraphics[width=1.0\textwidth]{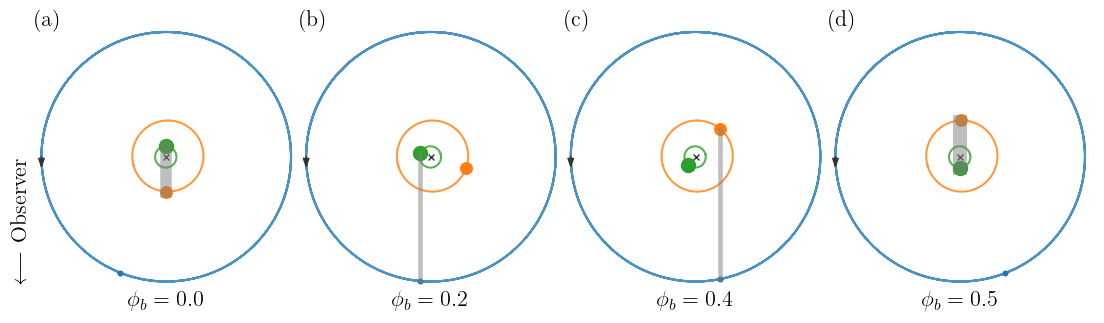}
\caption{Geometry of a CBP system (21 d G + M binary with a 92 d planet), where `x' marks the barycenter of the binary, and the green, orange, and blue symbols denote the position and orbit of the primary star, secondary star, and CBP, respectively. In this top-down view, the observer is located in 
the page, toward the direction indicated, such that the eclipsing geometry is met; the system is moving counterclockwise, as indicated by the arrow. Panels (a) through (d) depict consecutive snapshots of the system for \emph{one instance} of the 3-body orbit, during primary eclipse (PE), transit across the primary (TAP), transit across the secondary (TAS), and secondary eclipse (SE), respectively. The grey regions represent the shadow of the occluding body. The orbits are scaled to each other, but the relative sizes of the bodies are exaggerated for visual clarity. The conjunctions are indicated as a function of binary phase $\phi_{\mathrm{b}}$, since they depend upon binary parameters; however, transits will not always occur at those binary phases.}
\label{fig:cbp_geometry} 
\end{figure*}
A primary eclipse (PE), i.e., the deeper eclipse, occurs when the primary (hotter) stellar component is being eclipsed by the secondary; while a secondary eclipse (SE) occurs when the secondary is occulted by the primary. We abbreviate when the planet transits either the primary or secondary star as transit across the primary (TAP) or secondary (TAS), respectively. While occultations of a CBP by its binary hosts may occur, they are typically not detected in the system light curve, as the planet's brightness is usually negligible compared to the sum of flux output by the binary. Throughout the paper, we denote quantities associated with the binary and planet with subscripts `b' and `p', and use subscripts 1 and 2 to describe the primary and secondary stellar component, respectively. 

Figure~\ref{fig:cbp_geometry} shows a subset of eclipse/transit geometries for an illustrative CBP system, in which the planet orbits every 92 d around a 21 d G+M dwarf binary. The timing and duration of stellar eclipses and planetary transits depend on the relative positions and velocities between each pair of bodies. These are determined, to first order, by binary parameters, which set the binary orbit and stable configurations for the CBP orbit. 

In order for a CBP to be dynamically stable against perturbations by the binary, the planet must be located beyond a critical stability region. This critical semi-major axis $a_c$ was first quantified numerically by \cite{Holman1999} 
and subsequently confirmed and expounded upon by several authors (e.g., \citealp{Li2016, Quarles2018, Sutherland2019}). Typically, $a_c\approx(2-3)a_{\mathrm{b}}$, or equivalently, $P_c\approx(3-5)P_{\mathrm{b}}$. This inner stability limit sets the inner boundary to the CBP search period. 

\subsection{CBP Transit Detection}
\label{subsec:cbp_detection}

\subsubsection{Time-series Likelihood Ratio Test}
\label{subsec:likelihood_ratio}
Our detection algorithm relies on a model comparison technique, the likelihood ratio test. In essence, it compares the likelihood that a test CBP, corresponding to size $R_{\mathrm{p}}$ and period $P_{\mathrm{p}}$, transits one of the stellar components at every time, $t$, in the light curve data, relative to a null hypothesis that no transit is present. We quantify this ratio of likelihoods as the difference in $\chi^2$ between no-transit and transit, $\vec{\Delta \chi}^2  = \vec{\chi}^2_{\mathrm{null}} - \vec{\chi}^2_{\mathrm{CBP}}$. The \delchisq\ centered around cadence $i$ is given by
\begin{equation}
\label{eq:delchisq}
\begin{aligned}
\Delta \chi^2_i 
& = \sum_{i-N/2}^{i+N/2} \left( \frac{F_i - F_{\mathrm{CBP}, i}(R_{\mathrm{p}}\mathrm{=0}, P_{\mathrm{p}}, \vec{\theta_{\mathrm{b}}}) \cdot P_i(k)}{\sigma_{i}}  \right)^2\\
& - \sum_{i-N/2}^{i+N/2} \left( \frac{F_i - F_{\mathrm{CBP}, i}(R_{\mathrm{p}}, P_{\mathrm{p}}, \vec{\theta_{\mathrm{b}}}) \cdot P_i(k)}{\sigma_{i}}  \right)^2,
\end{aligned}
\end{equation}
where $\vec{F}$ and $\vec{\sigma}$ are the time-series data and associated uncertainties; $N+1$ is the length of the data segment being fit, which encompasses the duration of the test transit and continuum window surrounding it; $\vec{F}_{\mathrm{CBP}}$ is the model CBP transit light curve, which is a function of $R_{\mathrm{p}}$, $P_{\mathrm{p}}$, and binary parameters $\vec{\theta_{\mathrm{b}}}$; $\vec{P}(k)$ is a $k$th order polynomial, which is used to locally detrend the light curve. So, around each observed cadence $i$, we evaluate the difference in \chisq\ between a polynomial-only fit and a CBP transit centered at $i$ times a polynomial fit. We describe the CBP model which predicts $\vec{F}_{\mathrm{CBP}}$ in \S\ref{subsec:cbp_model}. 

This formalism assumes that every $\Delta\chi^2_i$ or $\vec{F}_{\mathrm{CBP}, i}$ samples the same number of data points $N+1$. As a result, any transit outside $N/2$ at the edges of a light curve will be missed. This problem is exacerbated in real observations, which may contain long gaps that separate the light curve into several data chunks. Therefore, in practice, we soften the requirement for strictly uniform $N+1$ sampling near edges, and adopt the reduced $\chi^2$ as a metric to evaluate the time-series likelihood ratios, to mitigate against edge effects.

When a transit is present at cadence $i$ (time $t$) in the data, the fit which includes the transit will give a lower $\chi^2_i$ than the fit without the transit, so $\Delta \chi^2_i$ will be positive when a transit is present, and conversely it will be negative when there is no transit. In evaluating the likelihood that a transit corresponding to a test CBP with period $P_{\mathrm{p}}$ and size $R_{\mathrm{p}}$ occurs at every observable cadence in the light curve, we construct a time-series \delchisq-curve. From here on, we drop the explicit arrow vector notation, and distinguish the vector (time-series) likelihood ratios as simply \delchisq\ and scalar (specified time) likelihood ratio as $\Delta \chi^2_i$. 


\subsubsection{EB Model}
\label{subsec:eb_model}
To remove eclipses (see panel 1b of Figure~\ref{fig:cbp_methods}) and predict transit depths, durations, and times during CBP-star conjunctions, we must determine the dimensions and orbital properties of the binary. We have previously characterized 728 well- to semi-detached \kepler\ EBs selected from the \kepler\ Eclipsing Binary  Catalogue\footnote{http://keplerebs.villanova.edu/} \citep{Prsa2011}. We refer the reader to \cite{Windemuth2019} for a detailed description and discussion of our EB modeling process. In the following, we present a brief summary of this method.

We combine the constraining power of light curves (LC) and spectral energy distributions (SED) of each binary, conditioned on stellar evolution models, to infer its absolute stellar (e.g., masses $M_1, M_2$ and \kepler-band flux ratio $F_2/F_1$) and orbital parameters (e.g., $P_{\mathrm{b}}, e_{\mathrm{b}}, i_{\mathrm{b}}, \omega_{\mathrm{b}}, t_{\mathrm{PE}}$). To solve this set of binary parameters $\vec{\theta_{\mathrm{b}}}$, the LC+SED forward model uses a Keplerian orbit solver coupled to a \cite{Mandel2002} analytic light curve model with quadratic limb darkening, and to the PARSEC \citep{Bressan2012} stellar isochrones. 

We use $\vec{\theta_{\mathrm{b}}}$ corresponding to the maximum-likelihood solution to predict the sky-projected binary positions as a function of time $t$, or binary phase $\phi_{\mathrm{b}}$. We consider the descending node of the binary orbit $\Omega$ to be $180^{\circ}$ along the $x$-axis, following the convention of \cite{Winn2010}. The $x$, $y$, and sky-projected positions $r_{\mathrm{sky}}$ of each binary component are then 

\begin{equation}
\label{eq:binary_positions}
\begin{aligned}
x_{1,2} & = -\mu_{1,2} r_{\mathrm{b}} \cos(\omega_{1,2}+f_{\mathrm{b}}),  \\
y_{1,2} & = -\mu_{1,2} r_{\mathrm{b}} \sin(\omega_{1,2}+f_{\mathrm{b}}) \cos i_{\mathrm{b}}, \\
r_{\mathrm{sky; 1,2}} & = -\mu_{1,2} r_{\mathrm{b}} \sqrt{1-\sin^2(\omega_{1,2}+f_{\mathrm{b}}) \sin^2 i_{\mathrm{b}}} \ ,
\end{aligned}
\end{equation}
where
\begin{equation}
\begin{aligned}
\mu_{1,2} &= \frac{M_{2,1}}{M_1+M_2},  \\
r_{\mathrm{b}} &= \frac{a_{\mathrm{b}} (1-e_{\mathrm{b}}^2)}{1+e_{\mathrm{b}}\cos f_{\mathrm{b}}}, \\
\omega_1 &= \omega_2 + \pi
\end{aligned}
\end{equation}
with $a_{\mathrm{b}}, e_{\mathrm{b}}, i_{\mathrm{b}}$ being the semi-major axis, eccentricity, and inclination of the binary, and $f_{\mathrm{b}} = f_{\mathrm{b}}(t)$ is the binary true anomaly.


\subsubsection{CBP Model \& Transit Regularization}
\label{subsec:cbp_model}
The CBP model uses the EB model (\S \ref{subsec:eb_model}) and assumes that the planet is a test body, i.e., $M_{\mathrm{p}}=0$, in a circular, edge-on Keplerian orbit around the binary barycenter. This approach makes simplifying assumptions to the 3-body problem. In reality, tidal or non-spherical distortions will alter the Keplerian binary orbit, and the changing binary potential modifies the CBP orbit. However, because the goal of this method is \emph{detection} rather than detailed characterization, our approximations are valid for their speed and accuracy, greatly reducing the size of parameter space which must be explored to make a detection (see also \S\ref{sec:discussion}). 

We do not know \emph{a priori} the period of the planet and the binary phase or time at which transits occur, except that the planet period must be greater than the critical period for stability $P_c$ \citep{Holman1999}. Therefore, we sample, i.e., test for the presence of a CBP transit, at every cadence in the observed light curve. Furthermore, we test the entire light curve for every trial $P_{\mathrm{p}}$ and $R_{\mathrm{p}}$, in a user-defined search grid. This is because the depth, duration, and timing of CBP transits depend on binary parameters $\vec{\theta_{\mathrm{b}}}$ as well as trial planet parameters. 

For a particular planet period and size, QATS-EB calculates the time-dependent positions of and relative separations between the planet and a stellar component \emph{at a proposed mid-transit time}. To regularize transits, we use the relative planet-star sizes and stellar flux ratio to determine the expected transit depth $\delta$, and use the time of first and fourth contact to evaluate the transit duration $\tau$ as a function of time $t$ or binary phase $\phi_{\mathrm{b}}$ (see panel 2b of Figure~\ref{fig:cbp_methods}). 

Additionally, we correct for the large-scale, binary-induced TTVs by realigning test transits across one of the binary components (a ``moving target") such that they occur at the (stationary) binary barycenter at $x=0$. We refer to the time between transit in front of a stellar component to transit in front of the barycenter as the barycentric time offset \tbary. This \tbary\ changes as a function of $\phi_{\mathrm{b}}$ and is approximately given by 
\begin{equation}
    t_{\mathrm{bary}} = \frac{x_{\mathrm{p}}(t_{\mathrm{tran}}) - x_{\mathrm{bary}}}{v_{x,\mathrm{p}}(t_{\mathrm{tran}})} = \frac{x_1(t_{\mathrm{tran}})}{v_{x,\mathrm{p}}(t_{\mathrm{tran}})} 
\end{equation}
which assumes that the planet is moving at constant velocity across the distance between the stellar component and binary barycenter, in the $x$-direction. Since the \delchisq-curve is a time-series likelihood ratio test for the presence of CBP transits, we apply this timing offset to all proposed mid-transit times, i.e., observed cadences. In doing so, we transform the entire \delchisq\ from observed times $t$ to $t^{\prime} = t-$\tbary.

Thus, for a given set of light curve data, binary parameters, and test CBP period and size, QATS-EB computes a regularized matrix $\Delta \chi^2 (P_{\mathrm{p}}, R_{\mathrm{p}}, t^{\prime})$. This timing transformation effectively reduces the observed, large-amplitude binary-induced TTVs.

\begin{figure*}
\includegraphics[width=1.0\textwidth]{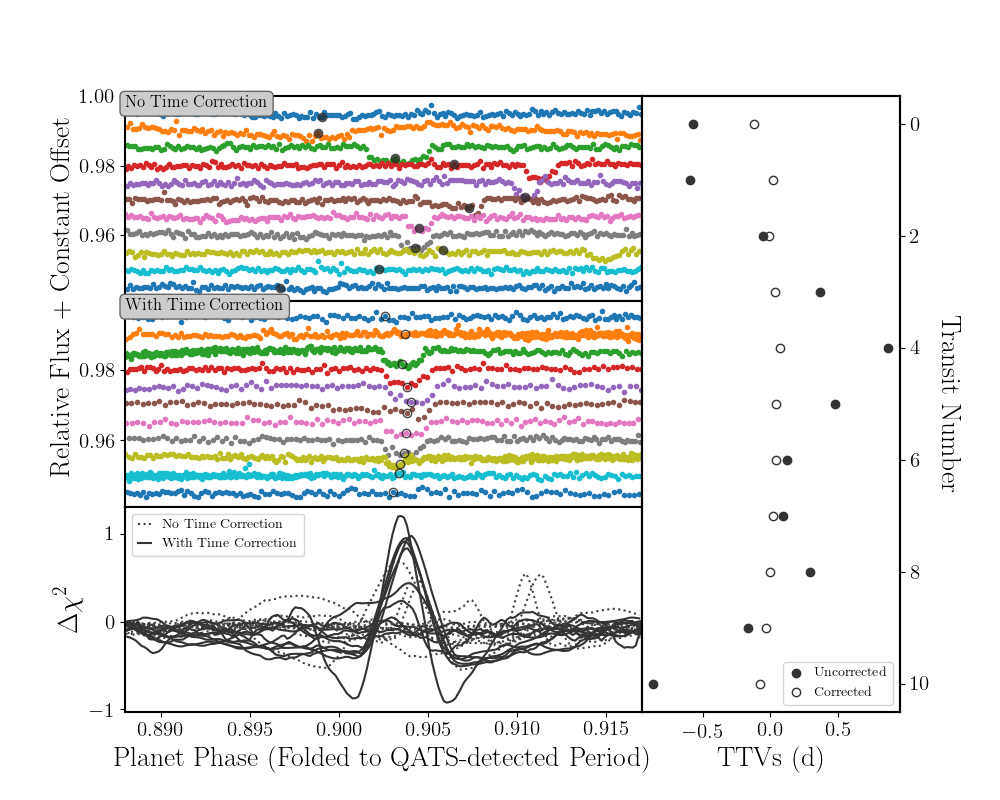}
\caption{Comparison of transits for \kepler-35\emph{b} in regular time $t$ and resampled time $t^{\prime}=t-$\tbary, where we apply QATS-EB with $f=0.02$ and $f=0.005$ to uncorrected and corrected cases, respectively. The left panels show the transits folded to their respective QATS-detected period, with arbitrary mean phase offset. Top panels show \kepler\ data, detrended with a linear polynomial and each transit number plotted with a constant flux offset, for visual effect. The dark circles denote the indices corresponding to the best QATS-detected period; note in the uncorrected case, the best detected period some times misses when true transits occur, due to their large TTVs. The bottom panel illustrates the corresponding expected \delchisq\ with the best QATS-detected transit depth and period. The transits are clearly more aligned when the barycentric timing offset is applied; the TTVs are reduced by an order of magnitude, from $\sim$0.5 d to $\sim$0.05 d (see right panel). }
\label{fig:cbp_tbary_kep35} 
\end{figure*}

\edit{To demonstrate the efficacy of the timing transformation, we show an example of its application to \kepler-35\emph{b} in} Figure~\ref{fig:cbp_tbary_kep35}. \edit{The figure} illustrates the differences in TTV amplitudes for \kepler-35\emph{b} with and without time correction. Here, we \edit{apply QATS-EB  with the same search parameters except} using $f=0.005$ and $f=0.02$ to $t^{\prime}$ and $t$, respectively, where $f$ is the fractional TTV window relative to the search $P_{\mathrm{p}}$ \edit{(see \S\ref{subsec:qats_significance} for further discussion of $f$)}. 
The transit light curves and corresponding \delchisq\ (left panels) are folded to their respective best detected periods, with the mean transit aligned to an arbitrary planet phase. For visual presentation, we vertically stack QATS-detected transit events and plot linearly detrended light curves, with dark circles denoting the QATS-detected times of transits. The left panels clearly demonstrate a significant reduction in TTVs when timing correction is applied; the mean TTV scatter decreases from $\sim$0.5 d to $\sim$0.05 d (see right panel). 

Finally, because CBP transit duration varies strongly as a function of binary phase (see panel 2b of Figure~\ref{fig:cbp_methods}), a \delchisq-curve  corresponding to a particular physical CBP model will also vary strongly as a function of binary phase. This is because for a CBP system, the majority of its light curve does not contain transits. Thus, in general, $\Delta \chi^2_i$ values are smaller for $\phi_{\mathrm{b}}$ or $t$ where the model predicts shorter CBP transits than for $\phi_{\mathrm{b}}$ with longer predicted transits, artificially causing the continuum of the \delchisq-curve to be sinusoidal. This quasi-periodic signal in the \delchisq\ continuum must be removed, as the standard QATS algorithm is not optimized for non-Gaussian noise.

To this end, and to make the search more computationally tractable, we adopt the following procedure in practice. First, we pre-compute a \delchisq$(\tau, \delta, t)$ matrix, i.e., in a 2D grid of duration $\tau$ and depth $\delta$, using a U-shape transit profile (see panel 2a of Figure~\ref{fig:cbp_methods}). \edit{Because the $\chi^2$ is quadratic as a function of depth, we can compute the \delchisq\ at any $\delta$ by evaluating the U-shape transit model at three grid points in $\delta$ \citep{Kruse2019}.} Then, we construct a CBP template, containing predicted transit depth, duration, and \tbary\ as a function of binary phase (see panel 2b of Figure~\ref{fig:cbp_methods}). Finally, we use that CBP template to linearly transform from the pre-computed \delchisq$(\delta, \tau, t)$ to the expected, or physical \delchisq$(P_p, R_p, t^{\prime})$ (see panel 2c of Figure~\ref{fig:cbp_methods}). 

In detail, we standardize the ``raw," pre-computed \delchisq$(\tau, \delta, t)$ matrix before using the CBP template to interpolate \delchisq\ as a function of physical parameters $R_{\mathrm{p}}$ and $P_{\mathrm{p}}$. Recall that at cadences without transits, $\Delta\chi^2_i$ changes strongly with duration. Thus, for each time-series slice of the pre-computed \delchisq$(\tau, \delta, t)$, larger duration values lead to more negative zero-point and higher scatter in the \delchisq\ continuum (see Figure~\ref{fig:cbp_methods} panel 2a), so that interpolating the raw \delchisq\ will introduce a variable continuum at the binary period. To avoid this correlated signal, we subtract each time-series \delchisq\ vector by its median, i.e., 50th percentile value and then normalize by the 68th percentile absolute difference value, a 1$\sigma$ estimate on the continuum scatter. 

Then, for a given test CBP $P_{\mathrm{p}}$ and $R_{\mathrm{p}}$, we linearly interpolate the expected $\Delta \chi^2_i$ value from the expected transit duration at each time $t$ and the standardized, pre-computed \delchisq\ grid. This removes binary phase dependence during interpolation and preserves the height of \delchisq\ ``pulses" -- when transits are actually present -- relative to the continuum (see panel 2c of Figure~\ref{fig:cbp_methods}). Finally, we transform to the barycentric time frame by resampling in uniform cadence $t^{\prime} = t-$\tbary.

\subsubsection{QATS Detection Significance}
\label{subsec:qats_significance}
We feed the regularized \delchisq$(P_{\mathrm{p}}, R_{\mathrm{p}}, t^{\prime})$-curves into standard QATS, which returns the maximum stacked transit signal $S_{\mathrm{best}}(P_p, R_p, f)$, where $P_p, R_p$ are the search period and size (depth), and $f$ is the user-specified, fractional transit timing variation window per $P_{\mathrm{p}}$. We build up a QATS quasi-periodogram or ``spectrum" (see panel 3a of Figure~\ref{fig:cbp_methods}) by searching through a range of planet periods, or, more precisely, range of minimum and maximum periods bounded by $f$. 

The choice of the quasi-periodic window is non-trivial, as a larger window is more likely to encompass transits with larger TTVs, but at the expense of including more background noise. \cite{Carter2013} recommended a constant fraction $f$ of the (average) search period, such that the difference between minimum and maximum quasi-periodic intervals scales linearly with the planet's orbit. This strategy is motivated by theoretical predictions of TTVs on a transiting CSP by additional perturbing bodies \citep{Agol2005}. In the CBP case, the large-scale, binary-induced TTVs scale with $(P_{\mathrm{b}}/P_{\mathrm{p}})^{2/3}$ (see, e.g., \citealp{Armstrong2013}). This motivates a choice for $f$ that increases more slowly with search period, which is attractive in that it includes less background noise at larger planet search periods. However, inaccuracies in the binary model and eccentricities in the planet's orbit will introduce additional timing error or variations, which can grow with planet period. For this paper, we adopt the fixed $f$ approach and find it effective in detecting the $P_{\mathrm{p}}<150$ d simulated and known \kepler\ CBPs. 

The continuum behaviour of the raw QATS spectrum depends on the noise properties of the light curve and corresponding \delchisq-curve and the input search parameters. To flatten the continuum, we fit a linear combination of polynomials in $P_{\mathrm{p}}$, the number of QATS-detected transits $N_{\mathrm{tran}}$, and the width of the transit timing variation window per search period, in cadences. We then subtract the linear combination model and normalize by the local QATS noise around each $P_{\mathrm{p}}$ (see panel 3b of Figure~\ref{fig:cbp_methods}; \citealp{Kruse2019}). This forms the final QATS ``signal-to-noise" and quantifies the detection significance.

\section{Example Applications}
\label{sec:application}
\subsection{Simulated Data}
\begin{figure}
\includegraphics[width=0.5\textwidth]{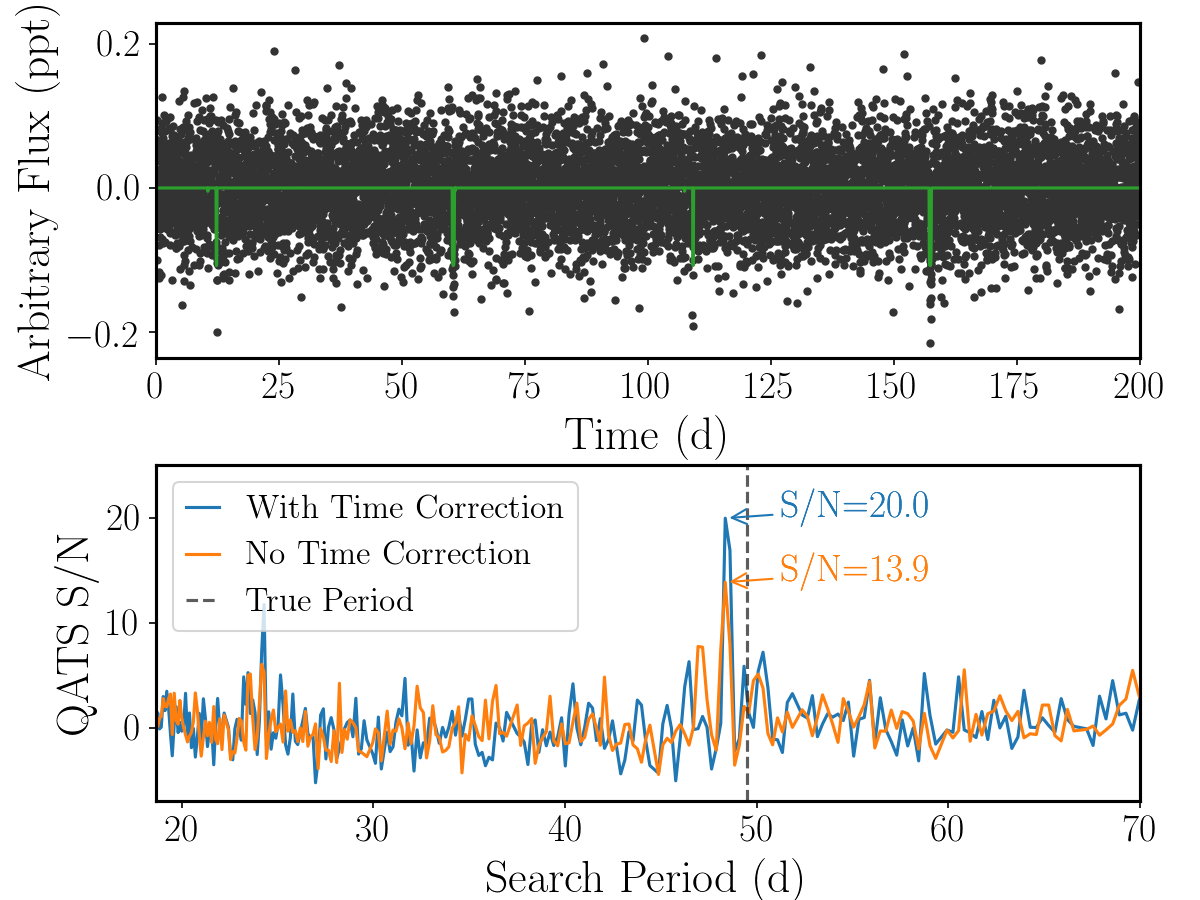}
\caption{Top: Simulated data (EB-removed) for a $1M_{\oplus}$), $1R_{\oplus}$) CBP around a 7.5 d G+M-dwarf binary with orbital and binary properties similar to \kepler-47 system \citep{Orosz2012-kep47, Orosz2019} (see Table~\ref{tab:parameters}). The simulated light curve, in arbitrary normalized, median-subtracted flux in parts-per-thousand (ppt), contains 4 transits across the primary ($\delta=0.0001$; the noiseless light curve is shown for reference). Transits across the secondary stellar component are not observed above the injected white noise threshold ($\sigma=5\times 10^{-5}$). Bottom: Recovered QATS signal-to-noise spectrum with a fractional TTV window $f$=0.01, i.e., 1\% of the search period. Blue and orange lines indicate QATS detection significance with and without barycentric timing offset correction. While the Earth-size planet is clearly detected in both instances, including timing correction yields a detection at the true period and with a significance that is 1.4 times higher than the uncorrected case.}

\label{fig:rebound} 
\end{figure}

We first test our technique on simulated data. We use the N-body integrator \texttt{rebound} \citep{Rein2012} to simulate a system with orbital and binary properties similar to \kepler-47\emph{b} \citep{Orosz2012-kep47}, but place an Earth-mass and Earth-size planet around the G and M dwarf binary. We integrate the system for 200 days (roughly 26.5 times the binary period), and tabulate the exact binary and planet parameter values used to initialize the simulation in Table~\ref{tab:parameters}. The sky-positions of each body are used to determine the system flux at every conjunction event via the \cite{Mandel2002} model. We inject white noise to the simulated light curve with 
$\sigma=5\times 10^{-5}$.
The top panel of Figure~\ref{fig:rebound} shows the EB-removed light curve of the simulated system. It contains 4 transits across the primary component ($\delta\approx0.0001$). Because the secondary star is much fainter than the primary ($F_2/F_1 \approx 0.006$), transits across the secondary are extremely shallow, below the level of injected noise $\sigma$. 

We apply QATS-EB to the light curve shown in the top panel of Figure~\ref{fig:rebound}, using binary parameter values listed in Table~\ref{tab:parameters} to compute expected \delchisq\ and a fractional TTV window $f=0.01$. That is, at each trial planet period, QATS computes the maximum total transit signal-to-noise with quasi-periodic bounding between successive transits which is 1\% of the search period wide. The resulting QATS detection significance is plotted in the bottom panel of Figure~\ref{fig:rebound}, where blue and orange colors indicate a QATS-EB search with and without timing corrections, respectively, as outlined in \S\ref{subsec:cbp_detection}. The Earth-size planet is detected at higher signal-to-noise ratio (S/N=20) when barycentric timing offsets are applied, about 1.4 times more significant than the non-regularized case (S/N=14).  

\begin{table*}[ht]
\small
\caption{Binary and Circumbinary Planet Parameters} \label{tab:parameters} 
\begin{tabular}{cccccccccccccccc}
\hline \hline
System    & $P_b$ & $e_b$ & $i_b$ & $M_1$ & $M_2$ & $R_1$ & $R_2$ & $F_2/F_1$ & $P_p$  & $e_p$ & $i_p$ & $R_p$ & $M_p$ \Tstrut \\ 
~ & (d) & ~ & $(^{\circ})$ & (M$_{\odot}$) & (M$_{\odot}$) & (R$_{\odot}$) & (R$_{\odot}$) & ~ & (d) & ~ & ($^{\circ}$) & (R$_{\oplus}$) & (M$_{\oplus}$) \Bstrut \\ \hline
Simulated* & 7.448 & 0.0288 & 89.34 & 1.04  & 0.36  & 0.964 & 0.351 & 0.00568   & 49.514 & 0.02  & 90    & 1     & 1    \Tstrut\Bstrut\\
Kep-35b$^{\dagger}$         & 20.733666      & 0.1421        & 90.4238        & 0.8877         & 0.8094         & 1.0284         & 0.7861         & 0.3941             & 131.458        & 0.04          & 90.76          & 8.16           & 40             \\ 
Kep-64b$^{\ddagger}$         & 20.000214      & 0.2117        & 87.36          & 1.384          & 0.387          & 1.75           & 0.422          & 0.00153            & 138.317        & 0.05       & 90.022         & 6.18           & $<$168         \Bstrut\\ \hline \hline 


\end{tabular}
\bigskip
Note: * modeled after \kepler-47b \citep{Orosz2012-kep47}; ${\dagger}$ \cite{Welsh2012}; ${\ddagger}$ \cite{Schwamb2013}. \Tstrut
\end{table*}

\subsection{Known \kepler\ CBPs}
Next, we demonstrate improved detection significance using QATS-EB on two known CBP systems, \kepler-35\emph{b} \citep{Welsh2012} and \kepler-64\emph{b}/PH1\emph{b} \citep{Kostov2013, Schwamb2013}. We show, for reference, the binary and CBP parameter values for both systems from their discovery papers in Table~\ref{tab:parameters}. Although both CBPs were discovered by eye, they were used as benchmarks for previous automated detection algorithms (\citealp{Klagyivik2017} with \kepler-35\emph{b}; \citealp{Armstrong2014} with \kepler-64\emph{b}) and provide a good baseline for comparison. 

We utilize \kepler\ simple aperture photometry (SAP) data, normalized by the median SAP flux of each observing quarter, spanning a total duration $\mathcal{T}\sim4$ years at a cadence of 0.0204 d (30 min). We subtract eclipse signals using the best-fit models from \cite{Windemuth2019}, and remove data points with quality flags $>$0, filling missing cadences with neighboring median flux values. We then apply QATS-EB to the ``cleaned" light curves. We employ a logarithmic depth grid of 1--30 times the median absolute difference in flux and a finely-spaced duration grid of 3--61 cadences (0.06--1.2 d) to construct the \delchisq$(\delta, \tau, t)$ matrix. We then transform this matrix to the expected \delchisq$(R_{\mathrm{p}}, P_{\mathrm{p}}, t^{\prime})$ using the best-fit binary parameters. We execute the quasi-periodic search with fractional TTV windows $f=0.005, 0.02$, initial and final planet periods $P_{\mathrm{p},i}=2.5P_{\mathrm{b}}$ and $P_{\mathrm{p},f}=\mathcal{T}/3$, i.e., requiring at least three primary transits. \edit{On a 2012 Intel$^{\textregistered}$ Core\texttrademark\ i5-3570 3.4GHz Linux machine with 4GB of memory, the QATS-EB search over $\sim$3x400 test CBP radii and periods takes $\sim$5 mins on each cleaned light curve ($\sim$65,000 cadences). This is approximately two orders of magnitude slower than standard QATS; however, the number of \kepler\ EBs around which to search for CBPs is roughly two orders of magnitude fewer than \kepler\ single stars. Thus, the total computation times are comparable. }

\begin{figure}
\includegraphics[width=0.5\textwidth]{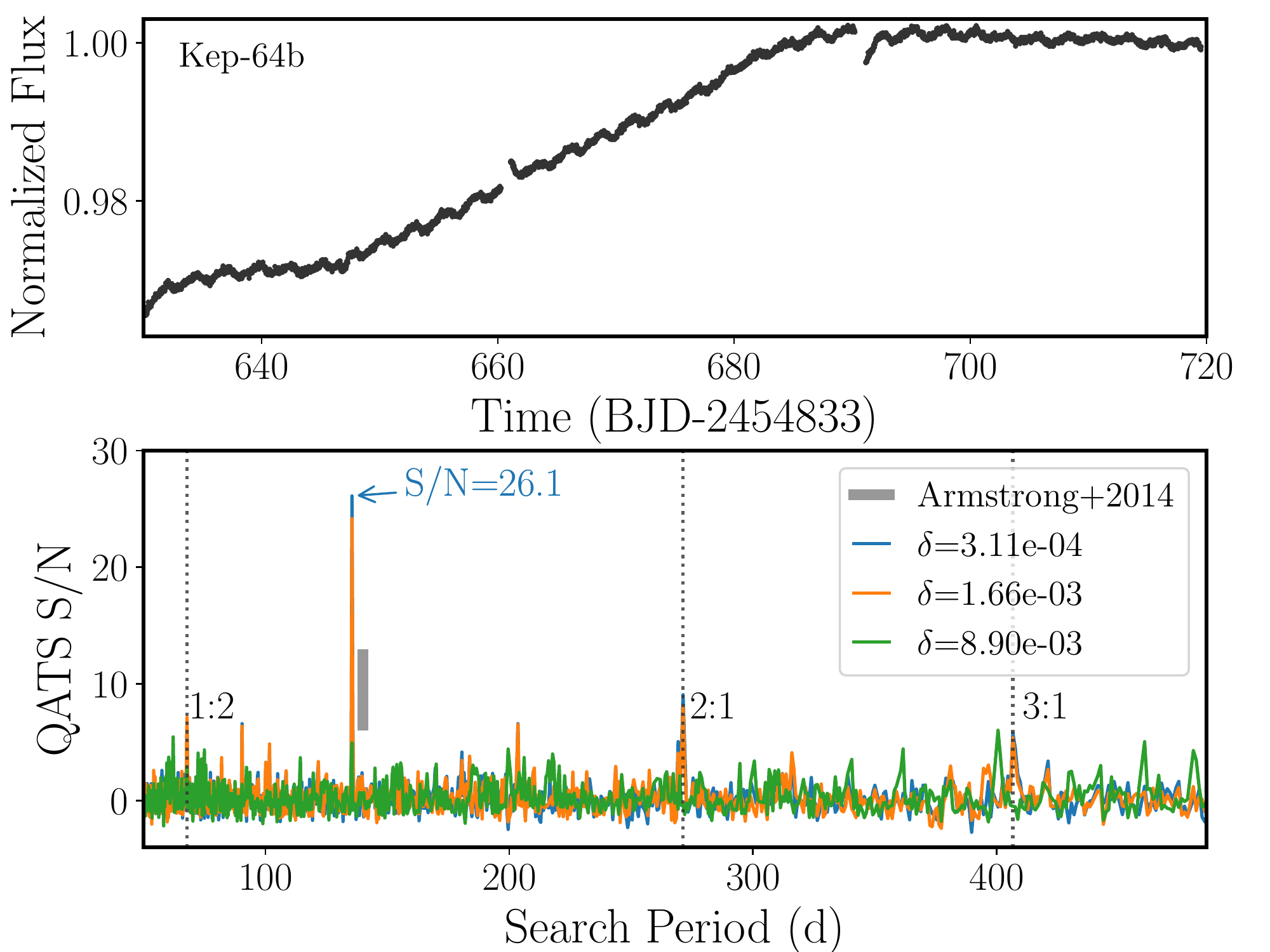}
\caption{Detection of \kepler-64\emph{b}. Top: a subset of the EB-removed SAP light curve, which shows high frequency stellar variability. Bottom: QATS-EB detection significance for different transit depths $\delta$ (CBP size). The grey region represent the relative detection significance from \citet{Armstrong2014}, based on estimated values from their Figure 1. Dotted lines mark P/2, 2P, 3P harmonics of the planet period. }
\label{fig:kep64b} 
\end{figure}

\edit{We show a subset of the EB-removed light curve of \kepler-64, on which we apply QATS-EB, and the resulting QATS detection spectrum with $f=0.005$ in the upper and lower panels of Figure~\ref{fig:kep64b}, respectively.} 
The QATS detection spectrum shows a strong peak (S/N=26) near the true period of 138 d with weaker peaks at P/2, 2P, and 3P aliases, at higher significance than \cite{Armstrong2014} values (grey region).  They defined detection significance as the mean value of the maximum peak and associated aliases, divided by the median periodogram value. This differs somewhat from the QATS S/N definition, which more closely resembles the Signal Detection Efficiency, defined as 
\begin{equation}
    \mathrm{SDE}= \frac{S_{\mathrm{peak}} - \mathrm{avg}(S)}{\mathrm{stdev}(S)} \ ,
\end{equation}
where $S$ is the signal \citep{Kovacs2002}. From Figure 1 of \cite{Armstrong2014}, we estimate an equivalent $\mathrm{SDE}_{\mathrm{hi}}$=(1.0-0.37)/0.05$\approx$12.5 and $\mathrm{SDE}_{\mathrm{lo}}$=(1.0-0.37)/0.1$\approx$6.5, which are 2 to 4 times smaller than the QATS-EB S/N. 

\begin{figure*}
\includegraphics[width=1.0\textwidth]{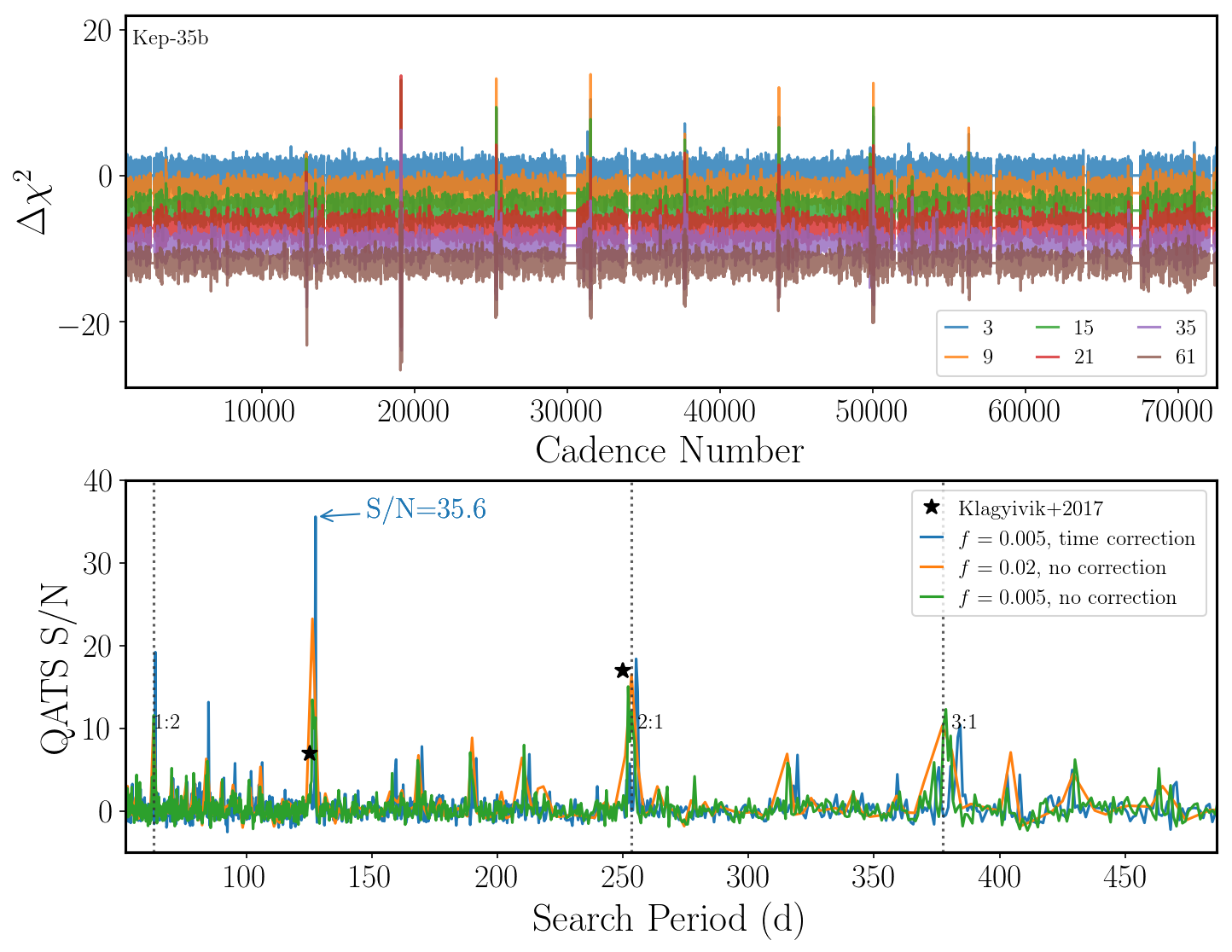}
\caption{Detection of \kepler-35\emph{b}. Top: \delchisq-curves as a function of test transit duration (in cadences), arbitrarily offset for visual effect, at a fixed transit depth ($\delta=4.58\times 10^{-4}$); the \delchisq($\delta, \tau, t$) matrix is used to compute the expected \delchisq($R_{\mathrm{p}}, P_{\mathrm{p}}, t^{\prime}$). Bottom: Resulting QATS-EB detection significance as a function of planet period in days, where different colors represent different search parameters and dotted lines denote aliases of the most likely period. Using the barycentric timing offset allows for smaller TTV windows $f$ in the QATS-EB search and significantly increases the detection S/N. With the timing correction, the planet is clearly detected near the true period of 131.5 d, at a S/N of 35.6. For comparison, \citet{Klagyivik2017} (shown as black stars) detected the 2:1 period alias at S/N=17 (comparable to QATS-EB with $f=0.02$ and no timing correction; orange line) and the true period at S/N=7. }
\label{fig:kep35b} 
\end{figure*}

Finally, we show the detection of \kepler-35\emph{b} in Figure~\ref{fig:kep35b}, and demonstrate how regularizing the transits using a physical model (2a--2c in Figure~\ref{fig:cbp_methods}) substantially improves the detection S/N. The top panel illustrates a slice in depth of the pre-computed \delchisq($\delta, \tau, t$) matrix, i.e., \delchisq\ as a function of time for different test transit durations and fixed depth $(\delta=4.58\times 10^{-4}$). The bottom panel plots the QATS S/N spectrum as a function of the quasi-periodic search period at fixed depth, under different search conditions.

Without the barycentric timing offsets to regularize transit times, QATS-EB requires a larger TTV window, i.e., includes more noise, for detection (see green vs. orange lines). Applying time corrections enables a reduction in $f$ and greatly increases detection significance (blue line). The CBP is clearly detected near the true period ($P_{\mathrm{p}}=131.5$ d) at S/N=35.6, with additional strong peaks at 1:2, 2:1, and 3:1 aliases of the period. This detection significance is 1.5 times higher than the case with $f=0.02$ and no timing correction. For comparison, we overplot detection results from \cite{Klagyivik2017} as black stars (see their Figure 1), where the highest peak from their analysis occurs near the 2:1 period alias, corresponding to a S/N of 17. They also recover the true period, although at lower detection significance (S/N$\approx$7). 

\section{Discussion}
\label{sec:discussion}
We have demonstrated the efficacy of QATS-EB to detect \edit{single} transiting CBPs in \S\ref{sec:application}. QATS-EB yields a 40\% increase in detection significance compared to standard QATS \citep{Carter2013}, and outperforms previous automated techniques applied to known \kepler\ CBPs \citep{Armstrong2014, Klagyivik2017} by factors of 2 to 4. 
Here, we revisit the underlying assumptions of QATS-EB and discuss their motivations and validity for this two-component technique.  

The original QATS algorithm maximizes the likelihood of transits in uniform cadence under the condition of quasi-periodicity and homogeneous transit duration. Applying this in its standard form to CBPs suffers from requiring a wide timing window, given their large, binary-induced TTVs, as well as from breaking the assumption of uniform transit duration. Instead, an ideal method for the detection of transiting circumbinary exoplanets would be to create a physical model for a CBP system, and to exhaustively search the parameter space of that model for the presence of a transiting planet (e.g., \citealp{Doyle1995}). However, this approach is extremely computationally expensive: given six CBP orbital elements and a radius relative to the size of the occulted stellar component, combined with the uncertainty in the number of CBPs in a system, this approach would require an initialization at an enormous number of models on a grid of instantaneous planet orbital periods, eccentricities, longitudes of periastron and ascending node, inclinations, phases, and radius ratios. 

QATS-EB is a compromise between these extremes:  we account for large-scale duration and timing variations using a simple, semi-analytic physical CBP model, while the flexibility of the standard QATS algorithm absorbs some of the inaccuracies incurred in this approximate model. The approximate CBP model, as detailed in \S\ref{sec:method}, operates under the following assumptions:


\begin{enumerate}
    \item It assumes that the binary parameters are perfectly known. This assumption can be thwarted by inaccuracies in the modeling of the eclipsing binary, and/or by difficulties with detrending the light curves.
    \item It assumes a circular orbit for the planet. Although the majority of known CBPs have small eccentricities ($\lesssim$0.05), the effect of eccentricity, when unaccounted for, may introduce error in transit shape and timing predictions.  
    \item It assumes that the planet's orbit is edge-on and that its impact parameter does not change significantly over the range of observed transits, i.e., that the depth of TAP and TAS is constant over the duration of observation. 
    \item It assumes that each transit event across one of the binary stars results in a single dip, rather than multiple dips during conjunction, which can occur when the star catches up and passes the planet multiple times.
    \edit{\item It assumes a single, test circumbinary planet. In principle, one can iteratively apply QATS-EB to detect additional planetary companions by masking transits associated with a detected candidate, as planet-planet perturbations are typically much smaller than binary-induced transit variations. However, multi-planet configurations with large transit timing and duration variations will require relaxing $f$ and reduce QATS S/N. }
\end{enumerate}
In the following subsections, we further discuss the scope of the limitations that \edit{assumptions 1--4} impose, as well as possible avenues for improvement. We argue that in most cases, the assumptions should not severely limit the performance of the search algorithm. We acknowledge that a more rigorous approach to quantify the algorithm's completeness \edit{and false alarm probability} requires applying QATS-EB en masse to a large grid of CBP simulations, \emph{with noise properties from or closely mimicking the data set upon which a transiting CBP search is applied}. This is beyond the scope of this current paper, and we leave performing a detailed injection and recovery analysis as a future task.
\subsection{Mass ratio and orbital eccentricity}
\begin{figure}
\includegraphics[width=0.5\textwidth]{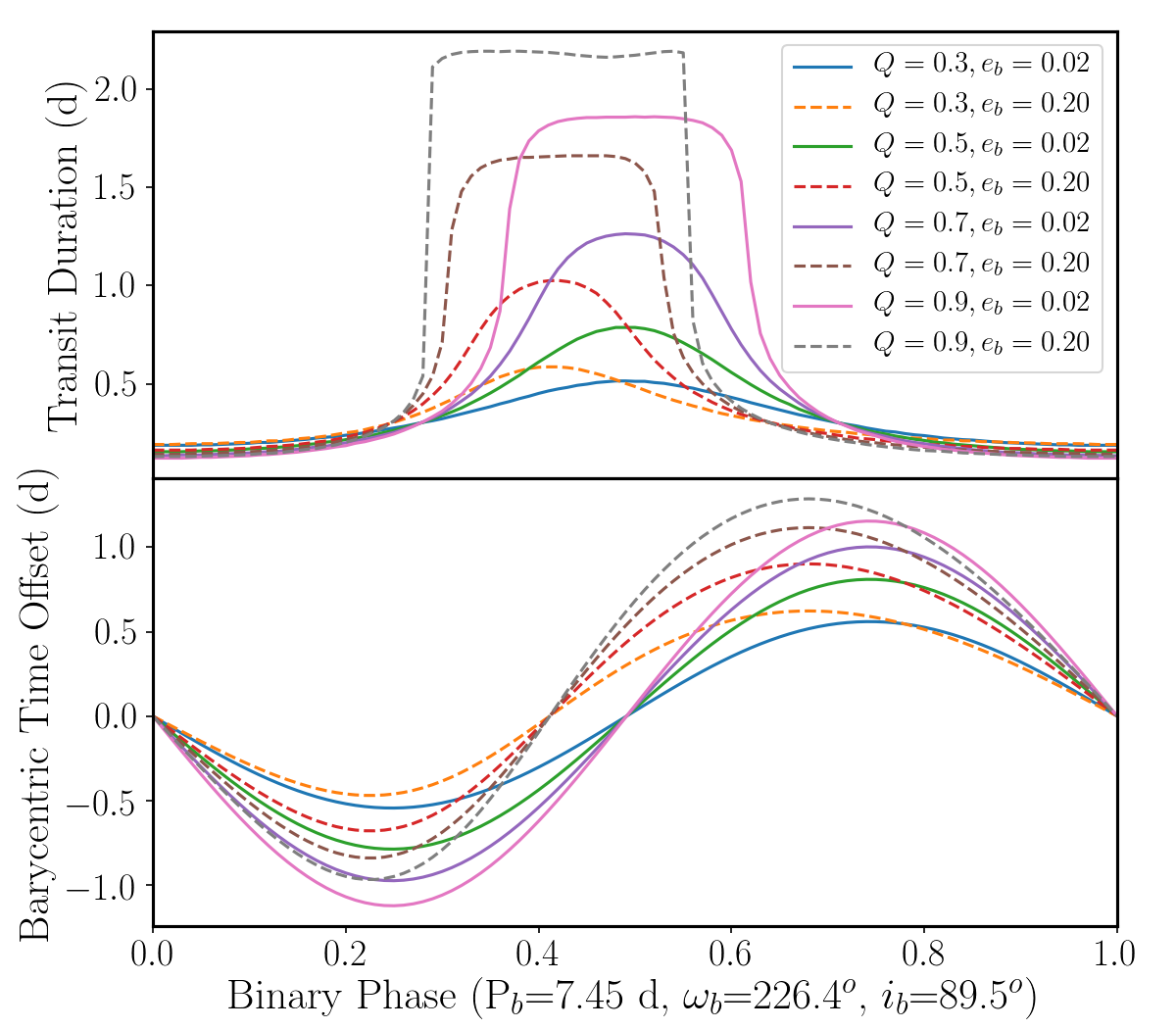}
\caption{The expected transit properties for an Earth-size CBP (P$_p$=60.9 d) in a circular, edge-on orbit around a 7.45 d binary as a function of binary mass ratio $Q=M_2/M_1$ and eccentricity $e_{\mathrm{b}}$. All other system parameters are held fixed. }
\label{fig:cbp_expected_transit_properties} 
\end{figure}

The standard QATS algorithm can mitigate the effects of assumptions \edit{1} (exact binary solution) and \edit{2} (circular planet orbit) when they are partly violated. If the mass ratio of the binary is inaccurate, this will affect the transit timing and duration amplitudes. Figure~\ref{fig:cbp_expected_transit_properties} shows the expected transit duration and timing offset as a function of binary phase, for a rocky CBP orbiting every 60.9 d around a 7.45 d binary with different mass ratios $Q=M_2/M_1$ and eccentricities $e_{\mathrm{b}}$. For reference, primary eclipse occurs at phase=0, while secondary eclipse occurs near phases of 0.5 and 0.4, for circular and eccentric binary cases, respectively. The expected transit duration and timing offset are largest for equal mass, eccentric binaries, in which binary reflex motion is greatest. Moreover, the greatest change in TAP duration occurs near phases surrounding secondary eclipse, when the sky-projected velocities of the primary star and planet are at their maximal amplitude and in the same direction. 

Likewise, deviations from a circular orbit of the planet will introduce differences in the transit timing and duration amplitudes relative to true values. We explore the effects of planet eccentricity by initializing \texttt{rebound} CBP models on a coarse grid of binary mass ratios $Q\in \{0.3, 0.9\}$, binary eccentricities $e_{\mathrm{b}} \in \{0.02, 0.2\}$, and planet longitudes of periastron $\omega_{\mathrm{p}} \in [0, 2\pi]$ in steps of 0.5, to populate transits in binary phase. We hold constant all other parameters: $P_{\mathrm{b}}$=7.45 d, $P_{\mathrm{p}}$=60.9 d, $\omega_{\mathrm{b}}$=226.4$^o$, $i_{\mathrm{b}}$=89.5$^o$, $i_{\mathrm{mutual}}$=0.2$^o$. We then compare QATS-EB predictions of transit duration ($\tau$) and barycentric time offset (\tbary) to simulation ``truths." 

\begin{figure}
\includegraphics[width=0.5\textwidth]{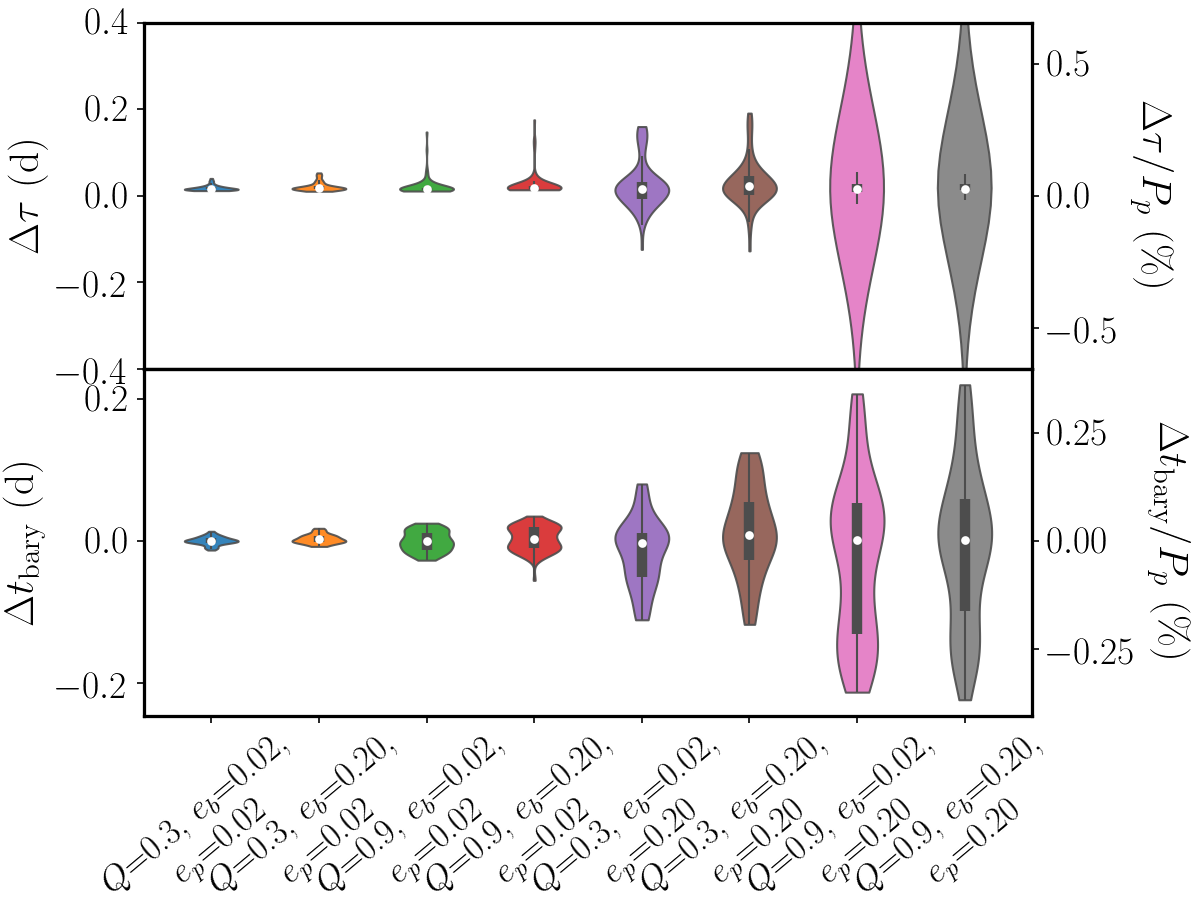}
\caption{The scatter distribution, marginalized over binary phase, between expected QATS-EB model values and simulation ``truths," for different CBP model initializations ($x$-axis). Top and bottom panels show differences in transit duration $\tau$ and barycentric timing offset \tbary\ in days (left axis) and per $P_{\mathrm{p}}=60.9$ d (right axis), respectively. The colored, shaded regions (``violins") represent the full distribution of difference values, where the widths indicate frequency of the dependent variable. The white circles correspond to second quartile (median) values; the thick dark box represents the inner quartile region (IQR), from first to third quartile range ($\pm$0.67$\sigma$). The thin dark lines or ``whiskers" denote the $\pm$1.5 IQR of the first and third quartiles ($\pm$2.7$\sigma$). Note that for $e_{\mathrm{p}}$=0.02 cases, the difference between true and QATS-EB values is $\sim$0.01 d, shorter than \kepler\ long cadence. For eccentric planets, QATS-EB predictions have larger error, as expected, with the bulk of scatter $\sim$0.1 d, about 0.2\% of the planet period.}
\label{fig:cbp_expected_transit_error}
\end{figure}

We show the scatter distributions of $\tau$ and \tbary, marginalized over binary phase, as ``violin" plots \citep{Hintze1998} in Figure~\ref{fig:cbp_expected_transit_error}. Here, the color regions (``violins") represent kernel density estimations of the scatter, i.e., $\Delta$\tbary\ and $\Delta \tau$ distributions, where the white circles denote median values. The dark boxes span first to third quartiles, or the inner quartile range (IQR; $\pm$0.67$\sigma$), and the whiskers span $\pm$1.5 IQR of the first and third quartiles ($\pm$2.7$\sigma$). For small planet eccentricities (see $e_{\mathrm{p}}$=0.02 cases), QATS-EB predictions typically agree with simulation values to within 0.01 d. For eccentric planets (see $e_{\mathrm{p}}$=0.2 cases), QATS-EB predictions typically differ by $\sim$0.05 d and $\sim$0.1 d -- roughly 0.1\% and 0.2\% of the planet period -- for low and high mass ratio binaries, respectively. The largest deviations in transit duration ($\gtrsim0.5$ d) occur near binary phases where the transit duration changes rapidly, i.e., when the projected $x$-velocity of the stellar component changes direction to match that of the planet's (see region around $\phi_{\mathrm{b}}$=0.3-0.7 in Figure~\ref{fig:cbp_expected_transit_properties}). Thus, even for modestly eccentric CBPs, applying \tbary\ reduces binary-induced TTVs by a factor of 5--10. 

The QATS algorithm can, in principle, correct for the error in the transit timing amplitude by using a wider user-specified window, but the variation in duration will lead to a slight loss in detection signal-to-noise. Fortunately, widening the QATS window causes the false-alarm floor to rise logarithmically, and so in many cases will not severely impact the transit detection capability of the algorithm
\citep{Carter2013}. For comparison, we tested QATS-EB on a circular ($e_{\mathrm{p}}=0.02$) vs. eccentric ($e_{\mathrm{p}}=0.2$) Earth-size CBP orbiting every 60 d around a $Q=0.9, e_{\mathrm{b}}=0.2, P_{\mathrm{b}}=7.5$ d EB, and found a S/N reduction factor of 1.5 for $f=0.01$.

\subsection{Time-varying CBP orbital elements}
QATS-EB assumes that the orbit of a CBP is edge-on when computing expected transit properties. For small deviations from this configuration, 
the change in expected $\tau$ and \tbary\ is small ($\lesssim$ 0.05 d). However, the orientation of a CBP and its mutual inclination with respect to the binary are not constant in time, even in stable configurations. Indeed, CBP orbits are perturbed by the time-varying binary potential, leading to precession and, in the case of eccentric binaries, libration of planet orbital elements \citep{Schneider1994, Farago2010, Doolin2011}; this oscillation of orbital elements can cause transits to disappear (e.g., \kepler-413\emph{b}; \citealp{Kostov2014}) or to appear (e.g., \kepler-47\emph{c}; \citealp{Orosz2019}) over time, as the system departs from and enters into a transiting geometry. In particular, the precession timescale for circular binaries is proportional to $P_{\mathrm{p}}$ and $(a_{\mathrm{p}}/a_{\mathrm{c}})^2$. For CBPs close to the stability region around short-period ($P_{\mathrm{b}}\approx10$ d) binaries, which describes the majority of the discovered population thus far, this timescale can be of order 1000s d, comparable to the length of observation for, e.g., \kepler. In such instances, the disappearance of transits during the total observing timespan, as is the case for \kepler-413\emph{b} \citep{Kostov2014}, leads to a detection signal-to-noise loss, as S/N $\propto \sqrt{N_{\mathrm{tran}}}$. 

Since it is highly inefficient for a detection algorithm to explore non-transiting parameter space, quantifying the limits of detection requires separately deriving geometric transit probabilities. This is non-trivial for CBPs, due to precession. For more details, we refer the reader to \cite{Li2016} and \cite{Martin2016}, who studied the transit probability of CBPs around circular binaries under finite observing times.


\subsection{Single vs. multiple transits during conjunction}
\label{subsec:single_transit}
Because the stellar components move along an orbit interior to a circumbinary planet, the differential relative speeds and sky positions may lead to multiple transits during one conjunction. Strictly speaking, there are two scenarios which result in multiple transits near one conjunction: (a) Near barycenter crossings, when the stellar components themselves are near eclipse, a CBP may consecutively transit one star then another; (b) When a CBP moves slowly relative to a stellar component, the star may pass through the planet's shadow multiple times, causing multiple transits across the same star \edit{(\citealp{Deeg1998}; see their Fig. 1)}. QATS-EB can, in principle, treat scenario (a), although in its current implementation it must search for transits across the primary independently of transits across the secondary, and vice versa. Scenario (b) forms a corollary to assumption \edit{4}, that a single transit or dip occurs each time the planet crosses one component of the binary. As we show below, this assumption, which we call the  ``single-transit-per-stellar-conjunction" or ``single-transit" criterion for short, is typically satisfied.

The single-transit criterion can be met if the speed of the planet is sufficiently large compared
to that of the primary (or secondary) star, such that the star does not have time to loop around and intersect the planet's shadow along the line-of-sight multiple times. We
explore in the appendix the range of parameters for which there will be a single vs. multiple transits
of the primary and secondary stellar component by the planet when it crosses the
binary, in the limit of circular, edge-on orbits for both planet and binary. 
We derive the critical planet velocity for single-transits marginalized over binary phase and compute the fraction of time corresponding to single-transits of the primary and secondary (see appendix). We show that for CBP systems with semi-major axis ratios $a_{\mathrm{p}}/a_{\mathrm{b}}$ ranging from 3 to 10 and binary star mass ratios $Q=M_2/M_1$ ranging from zero to one, the probability of a single-transit exceeds 86\% for both the primary and secondary star in the circular edge-on case.  We expect that for modest binary eccentricities this will also hold true. The single-transit constraint required by QATS-EB, then, applies to the bulk of parameter search space, except for pathological cases.

\section{Conclusions}
\label{sec:conclusions}
We have developed an automated transit detection algorithm tailored to find circumbinary planets, which exhibit large transit timing, duration, and depth variations due to reflex motion of binary hosts. Our technique, dubbed ``QATS-EB," uses a physical EB+CBP model to regularize the transits with respect to the binary barycenter, and constructs a time-series in CBP transit likelihood. QATS-EB permits additional TTVs after regularization, i.e., due to model inaccuracies, by using the quasi-periodic search algorithm developed by \cite{Carter2013}. This two-component technique enhances overall detection significance, which is particularly important when searching for CBPs that are small and/or exhibit sparse transits. Figure~\ref{fig:cbp_methods} shows a summary of the search procedure and detection statistic. 

To balance computational efficiency and model complexity, the semi-analytic CBP model component of our technique makes several simplifying assumptions. Namely, it is conditioned on edge-on, circular planet orbits. For modest eccentricities, typical errors in regularization of transit timing and duration are small ($\lesssim$0.05 d). Moreover, the quasi-periodic nature of the standard QATS algorithm is able to alleviate larger timing deviations with a wider user-specified window, although doing so encompasses more background noise. 

As demonstrated for \kepler-35 and \kepler-64 systems, our method significantly improves detection signal-to-noise ratio, by factors $>$2. This improvement boosts our chances to detect smaller circumbinary planets, e.g., in the terrestrial regime. We plan to use our technique for a blind CBP search around \kepler\ EBs and test detection completeness to infer CBP population statistics in a following paper. Although we intend to search for CBPs in \kepler\ data, our technique can be applied to other time-series data with uniform cadence and sufficient monitoring baseline, such as those from \emph{K2}, the continuous viewing zones of \emph{TESS}, or the upcoming PLATO mission.

\section*{Acknowledgements}
This work was supported by National Aeronautics and Space Administration (NASA) Headquarters under the NASA Earth and Space Science Fellowship Program Grant NNX15AT44H, and by NASA Astrophysical Data Analysis Program Grant NNX13AF20G.
D.W. acknowledges support from the Virtual Planet Laboratory and the Astrobiology Program at the University of Washington. 
E.B.F. acknowledges support from the Penn State Eberly College of Science and Department of Astronomy \& Astrophysics, the Center for Exoplanets and Habitable Worlds and the Center for Astrostatistics.  
E.B.F. acknowledges support during residency at the Research Group on Big Data and Planets at the Israel Institute for Advanced Studies.  \edit{N.H. acknowledges support from the NASA XRP Program Grant 80NSSC18K0519.} 
\edit{We thank the referee P\'eter Klagyivik for their thoughtful review and useful suggestions.}
We acknowledge many valuable contributions with members of the \kepler\ Science Team's working groups on multiple body system, transit timing variations, and eclipsing binary working groups.  
We thank the entire \kepler\ team for years of work leading to a successful mission and data products critical to this study.  
%
%


\bibliographystyle{mnras}
\bibliography{paper}

\clearpage

\appendix
\section*{Appendix}
\label{appendix}
As discussed in \S\ref{sec:discussion} and \S \ref{subsec:single_transit}, the QATS-EB algorithm assumes
only a single planetary transit across one of the stellar components will occur at each passage of the planet across the
binary. We refer to this as the ``single-transit-per-stellar-conjunction" -- or the abbreviated ``single-transit" -- criterion. For some binary and planet 
configurations, multiple transits across the same star can occur as the planet crosses the line-of-sight between observer and binary component \edit{\citep{Deeg1998}}.   

\begin{figure}
    \centering
    \includegraphics[width=0.45\textwidth]{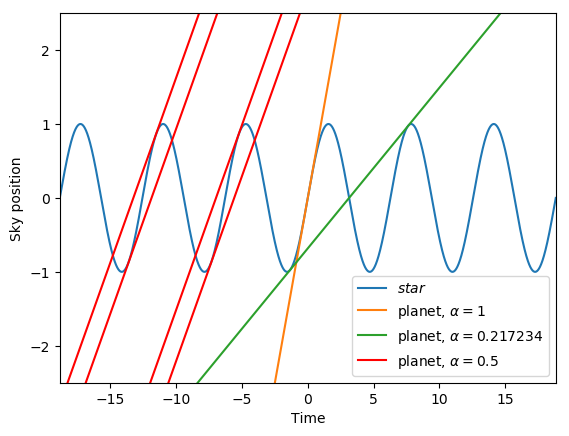}
    \caption{Sky position versus time of one component of
    the binary star (blue) and planets with various speeds.
    The $x$-axis has units in which the binary orbital period is $2\pi$,
    while the $y$-axis is in units of the amplitude of the binary star's
    orbit.  The speed of the planet's orbit, $\alpha$, is in units of
    the speed of the binary star's orbit.}
    \label{fig:spacetime}
\end{figure}

In this appendix, we estimate the fraction of time that this single-transit criterion is met, in the approximation that both the binary and planet are on circular, edge-on
orbits.  Here, we approximate the planet's motion as linear, as its orbital frequency is at least $\sim$three times slower than that of the binary, for stability \citep{Holman1999}. Figure \ref{fig:spacetime} shows the sky-projected orbital
position as a function of time for one of the binary components and the planet \edit{(see also Figure 1 of \citealp{Deeg1998})}. For
only one transit to occur, the planet's path 
must only intersect the sky position of each binary star once.

To translate this constraint into physical parameters requires comparing the orbital speed of the planet, $v_{\mathrm{p}}$, to the orbital speed of each binary star, $v_{1,2}$. We define the ratio of maximal velocities between the planet and one of the stellar components as 
\begin{equation}
\label{eq:alpha}
    \alpha_{1,2} = \left(\frac{a_{\mathrm{p}}}{P_{\mathrm{p}}}\right) \left(\frac{P_{\mathrm{b}}}{a_{\mathrm{1,2}}}\right)
\end{equation}
This $\alpha$ value must be sufficiently large, i.e., steep as a function of binary orbital phase, so that it only intersects the binary star once.

In Figure~\ref{fig:spacetime}, we show three examples of the relative planetary speed $\alpha$ (here, we omit specification of primary or secondary star and thus drop the subscripts 1 and 2) . For $\alpha=1$, the planet's speed matches the maximum speed of the binary star, and thus the planet will only cross the star's path once at any orbital phase.  For $\alpha = 0.217234$, the planet is
moving slowly enough that it will always intersect the star's position
three times as it crosses the binary.  For intermediate values of
$\alpha$ there will be either one or three transits as the star
crosses the binary.  An example is shown on the left with $\alpha=1/2$
for which there are three transits in between the closely spaced
phases, and one transit elsewhere.

\begin{figure}
    \centering
    \includegraphics[width=0.45\textwidth]{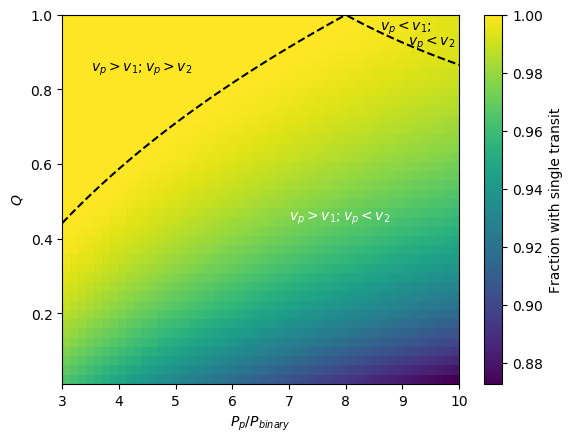}
    \caption{
    Fraction of phases for either star, $(\phi(m_1)+\phi(m_2))/2$, at which a transiting planet
    will have a single transit for both stars.  The range of binary mass-ratio, $Q$, and period ratio,
    $P_p/P_{\rm binary}$, for which the planet speed exceeds the stellar
    speed assuming circular, edge-on orbits is indicated with the black dashed lines.}
    \label{fig:planet_speed}
\end{figure}
We now consider the fraction of binary phases that satisfy the single-transit criterion for \emph{both} stellar components of the binary. Even if the velocity constraint $v_{\mathrm{p}}\ge v_{1,2}$ is violated, the QATS-EB algorithm 
may still apply, as the assumption of single transits is still valid at certain ranges in binary phase.
Figure \ref{fig:planet_speed} shows the fraction of binary phases that meet the single-transit criterion for both stars.  From Equation \ref{eq:alpha} and defining $Q=M_2/M_1$, we derive
\begin{eqnarray}
    \alpha_1 &=& \frac{1+Q}{Q} \left(\frac{P_{\rm p}}{P_{\rm b}}\right)^{-1/3}\\
    \alpha_2 &=& (1+Q) \left(\frac{P_{\rm p}}{P_{\rm b}}\right)^{-1/3},
\end{eqnarray}
This implies, as Figure~\ref{fig:spacetime} illustrates, that one transit per stellar conjunction occurs at all phases for $\alpha_1, \alpha_2 >1$.
When $\alpha_1, \alpha_2 < \alpha_{\mathrm{crit}}$, where $\alpha_{\mathrm{crit}} = 0.217234$, then there
are 3 or more transits.  For $\alpha_{\mathrm{crit}} < \alpha_1, \alpha_2 < 1$, the
fraction of binary phase (time) during which there is one transit is given by:
\begin{equation}
    \phi_{\mathrm{b}}(\alpha) = \begin{cases}
    0 & \alpha \le \alpha_{\mathrm{crit}}\\
    1-\frac{1}{\pi}\left(\frac{\sqrt{1-\alpha^2}}{\alpha} - \cos^{-1}(\alpha)\right) & \alpha_{\mathrm{crit}} < \alpha \le 1\\
    1 & \alpha > 1
    \end{cases}
\end{equation}
The function $(\phi_{\mathrm{b}}(\alpha_1)+\phi_{\mathrm{b}}(\alpha_2))/2$ is plotted in Figure \ref{fig:planet_speed}.
Thus, across a wide swath of parameter space in $Q$ and $P_{\mathrm{p}}/P_{\mathrm{b}}$, the single-transit-per-stellar-conjunction criterion is satisfied for most eclipsing binary phases, under the assumption
of circular orbits for the binary and the planet.



\bsp	
\label{lastpage}
\end{document}